\documentclass[12pt]{article}
\usepackage{graphicx}

\newcommand\pubnumber{SNSN-323-63}
\newcommand\pubdate{\today}

\def\pisa{Department of Physics, University of Pisa\\
and INFN Pisa, I-56127 Pisa, ITALY}

\def\Title#1{\begin{center} {\Large #1 } \end{center}}
\def\Author#1{\begin{center}{ \sc #1} \end{center}}
\def\Address#1{\begin{center}{ \it #1} \end{center}}

\newcommand\pubblock{\rightline{\begin{tabular}{l} \pubnumber\\
         \pubdate  \end{tabular}}}
\newenvironment{Abstract}{\begin{quotation}  }{\end{quotation}}
\newenvironment{Presented}{\begin{quotation} 
      \begin{center}\begin{large}}{\end{large}\end{center} \end{quotation}}

\def\etal{\emph{et al.}}

\def\beq{\begin{equation}}
\def\eeq#1{\label{#1}\end{equation}}
\def\eeqn{\end{equation}}

\def\beqa{\begin{eqnarray}}
\def\eeqa#1{\label{#1}\end{eqnarray}}
\def\eeqan{\end{eqnarray}}

\let\bar=\overbar

\def\etal{{\it et al.}}

\def\Dslash{\not{\hbox{\kern-4pt $D$}}}
\def\dslash{\not{\hbox{\kern-2pt $\del$}}}

\def\msb{{\bar{\ssstyle M \kern -1pt S}}}

\begin{document}
\begin{titlepage}
\pubblock

\vfill
\Title{Current and future kaon experiments}
\vfill
\Author{Marco S. Sozzi}
\Address{\pisa}
\vfill

\begin{Abstract}
An overview of the recent past, present and future of flavour physics 
studies using kaon decays is presented.
\end{Abstract}
\vfill

\begin{Presented}
Proceedings of CKM2010, the 6th International Workshop 
on the CKM Unitarity Triangle, University of Warwick, 
UK, 6-10 September 2010
\end{Presented}
\vfill

\end{titlepage}

\def\thefootnote{\fnsymbol{footnote}}
\setcounter{footnote}{0}

\begin{flushright}
\emph{Dedicated to the memory of N. Cabibbo (1935-2010).}
\end{flushright}

\vspace{-0.5cm}
\section{Mandatory introduction}

Kaons have been a remarkably effective tool to provide insight for 
shaping the Standard Model (SM) as we know it. 
Besides originating the introduction of the very concept of flavour, the kaon 
system played the key role in the discovery of such phenomena as parity and 
CP violation, the GIM mechanism and the existence of charm, and was central 
in the investigation of lepton flavour and CPT symmetries. 
Besides historical reasons, this richness of results is partly due to this 
system being a rather simple one, while exhibiting all the subtleties linked 
to flavour physics. \\
Sixty-six years since the first published observation of a K meson in cosmic rays 
one might be excused for wondering whether the information 
which these particles can provide on Nature has been eventually exhausted. 
We anticipate that the answer is negative. \\

\vspace{-0.5cm}
\section{Glorious past}

The copious amount of results on kaon decays obtained since 
the late '90s originates from the intense experimental effort aimed at the measurement 
of direct CP violation, which led to three experiments being built at CERN (NA48), 
Fermilab (KTeV) and Frascati (KLOE).
As is well known, such experimental endeavour led to the proof of the existence 
of direct CP violation through the measurement of the $\epsilon'/\epsilon$ parameter \cite{eprimeNA48} \cite{eprimeKTeV}, just shortly before B factories started 
measuring such phenomenon in B meson decays. 

The importance of the $\epsilon'/\epsilon$ measurement is qualitative, 
and lies in being the first verification of the CKM flavour picture of the 
Standard Model, proving that CP violation is indeed a ubiquitous property of 
weak interactions (as B factories impressively showed). 
Nevertheless, with the recent final analysis of KTeV data \cite{eprimeKTeV}, 
$\epsilon'/\epsilon$ is measured with rather good (12\%) precision; unfortunately, 
so far, such precision could not be exploited quantitatively to constrain in a significative way the Standard Model through the ``unitarity triangle'', due to 
dire theoretical difficulties.
A somewhat better situation exists for the oldest (indirect) CP violation 
parameter $\epsilon$, which is now measured to a very good precision (below 0.5\%), 
but whose potential in constraining the SM is also reduced by the theoretical 
difficulties related to low-energy hadron physics. \\
High hopes for theoretical improvements on these points lie in the steady progress 
of lattice QCD, which has the potential of transforming existing K physics 
measurements of CP violation in additional significant constraints on the closure 
of the unitarity triangle.

The legacy of CP-violation K experiments is however much larger than what was 
mentioned above. 
Besides leading to the development of state-of-the-art detectors and 
innovative detection and analysis techniques, many more physics results 
than initially anticipated were obtained as byproducts of the above 
investigations.
Just to quote a couple among many: possible large CP violating effects driven by 
supersymmetry were excluded by NA48 with much improved limits on Dalitz plot 
asymmetries in $K^\pm \rightarrow 3\pi$ decays \cite{K3pi}, and stringent limits 
on several flavour-violating decays were obtained by KTeV (e.g. \cite{LFVKTeV}).

While NA48 and KTeV are hadron beam fixed-target experiments, KLOE, being a 
Kaon factory exploiting coherent pair production in $e^+ e^-$ resonant 
production, presents quite different strengths and weaknesses; in particular 
it has the capability of exploiting its absolute flux measurement for 
BR measurements, and is a unique facility for the production of tagged $K_S$ 
particles \cite{KLOE}. Among the many results obtained, we highlight the 
recent study of semi-leptonic decays of $K_S$, with the first measurement of 
the (indirect) CP charge asymmetry in $K_S$ \cite{KS_KLOE}: while still well 
compatible with zero at present, such measurement is a first step towards a new 
CPT symmetry test by comparison with the corresponding $K_L$ asymmetry, 
measured with 2\% precision.

As another byproduct of the above experimental activities, much improved 
measurements of kaon Branching Ratios (BR) were obtained.
The increased care in handling systematic issues, as required by the larger 
statistics available, led to an important shift in the value of most BRs, 
starting from the comprehensive set of relative measurements of $K_L$ BRs 
by the KTeV experiment \cite{KTeV_BR}.
Further precise data obtained by NA48 \cite{NA48_BR} and KLOE \cite{KLOE_BR} 
\cite{KLOE_BR2}, also on $K^\pm$ and $K_S$ (and also BNL E865 \cite{E865_BR}), 
eventually led to the present state of kowledge, significantly different with 
respect to older PDG compilations. 
The single most significant issue responsible for the shifts was the proper 
handling of radiative corrections, quite important at the present level of 
statistical accuracy, and sometimes not correctly treated in older data.

\vspace{-0.5cm}
\section{Intriguing present}

The campaign of neutral and charged kaon BR measurements was partly driven by 
the interest in checking some ``tension'' between the data and the unitarity 
constraint for the first row of the CKM matrix: flanking improved measurements 
of $|V_{ud}|$ from beta decays, the knowledge of the value of $|V_{us}|$ 
(Cabibbo angle) from semileptonic kaon decays $K \rightarrow \pi \ell \nu$
($K_{\ell 3}$) was significantly improved.
From the new set of BR measurements, and the critical re-examination of the older 
ones and their systematic error correlations, the experimental accuracy on the 
determination of $|V_{us}|$ is now close to 0.2\%, with full statistical 
consistency between individual measurements. 
A detailed analysis of the above measurements can be found in \cite{Vus}. \\
Actually, $|V_{us}|$ cannot be extracted from the data without using 
theoretical input on the normalization of the $K \rightarrow \pi$ vector form factor 
at zero momentum transfer. Such normalization is nowadays obtained from unquenched 
lattice QCD simulations \cite{Lattice2010} with about 0.5\% precision and a 
consistency around 1\%.
Theoretical precision is thus approaching the experimental one, further improving 
the embarassingly good agreement of data with the SM.
The comparison of $|V_{us}|$ measurements from different decay modes also allows 
a test of lepton universality, passed with flying colours at 0.5\% precision
\cite{Vus}. 

Related measurements involve charged kaon leptonic decays $K \rightarrow \ell \nu$ 
($K_{\ell 2}$).
A renewed interest in such decays first arose from the observation 
\cite{Marciano} that improved Lattice QCD measurements of the ratio of kaon and 
pion decay constants ($f_K/f_\pi$) allow extracting $|V_{us}|$ from precise 
measurements of the $K_{\mu 2}$ BR.
The lattice precision on $f_K/f_\pi$ is now approaching the 1\% level 
\cite{Lattice2010}, while the experimental accuracy on the $K_{\mu 2}$ BR is 
below 0.3\% \cite{KLOE_BR2}.

It is also worth stressing that leptonic and semi-leptonic decays of pseudo-scalar 
mesons probe possible New Physics (NP) contributions with different chirality 
structures, so that the comparison of the two classes of decay modes (respectively 
helicity suppressed and allowed) allows placing bounds on some non-SM amplitudes 
in a given model: in the minimal supersymmetric extension of the SM such bounds relate 
to the value of the charged Higgs mass $m_{H^\pm}$ and the $\tan \beta$ parameter, 
excluding regions at low $m_{H^\pm}$ and large $\tan \beta$ \cite{Vus}.

It was also noted \cite{Masiero} that lepton-flavour violating terms, which  
appear in SUSY extensions of the SM, might affect the ratio of $K^\pm_{e2}$ to 
$K^\pm_{\mu 2}$ BRs (known in the SM to 0.04\% precision) up to the percent level 
in some regions of parameter space.
A recent measurement by KLOE \cite{KLOE_Ke2} improved the experimental knowledge 
of the above ratio to 1.2\%.
A dedicated measurement of such universality in charged kaon leptonic decays was 
undertaken by the NA62 collaboration at CERN, using the NA48/2 apparatus. The main 
experimental difficulty in this measurement lies in the control of the 
background to $K_{e2}$ from $K_{\mu 2}$ decays in which a muon radiates a 
very hard photon. The result from the analysis of 40\% of the collected data 
\cite{NA62_Ke2} has 0.5\% precision and is consistent with lepton 
universality and the SM.

\vspace{-0.5cm}
\section{Challenging future}

As the great amount of flavour data collected in recent years shows, New Physics 
flavour effects cannot be large.
After the general verification of the overall flavour structure of the SM, 
the trend towards finer and more stringent quantitative SM tests, looking for 
deviations in a search for NP, finds once more kaons in the first line. 

It has been long known that a handful of ultra-rare kaon decay modes 
can act as very powerful probes of flavour physics \cite{rare}, due to some unique 
features resulting in a few percent intrinsic precision in the theoretical 
BR prediction, an almost unique case in hadronic physics.
Some features of these modes are summarized in table \ref{tab:rare}; the 
hard-GIM suppression in the SM allows potentially large NP contributions, and 
the extraction of the hadronic part of the amplitude from that of 
$K_{\ell 3}$, experimentally very well known, are the two key points to understand 
the relevance of these decays.

\begin{table}[hbt]
\begin{center}
\begin{tabular}{|l|c|c|}
\hline
Mode & BR(SM) & BR(exp) \\
\hline
$K_L \rightarrow \pi^0 e^+ e^-$ & 
  $10^{-11}$ ($3 \cdot 10^{-12}$ CPV$_{\mathrm{dir}}$) & $< 2.8 \cdot 10^{-10}$ \\
$K_L \rightarrow \pi^0 \mu^+ \mu^-$ & 
  $10^{-11}$ ($1 \cdot 10^{-12}$ CPV$_{\mathrm{dir}}$)& $< 3.8 \cdot 10^{-10}$ \\
$K^+ \rightarrow \pi^+ \nu \overline{\nu}$ & 
  $(8.5 \pm 0.7) \cdot 10^{-11}$ & $1.47^{+1.30}_{-0.89} \cdot 10^{-10}$ \\ 
$K_L \rightarrow \pi^0 \nu \overline{\nu}$ & 
$(2.6 \pm 0.4) \cdot 10^{-11}$ & $< 2.6 \cdot 10^{-8}$ \\
\hline
\end{tabular}
\caption{Ultra-rare kaon decay modes. CPV$_{\mathrm{dir}}$ denotes the 
short-distance, direct-CP violating part of the BR.}
\label{tab:rare}
\end{center}
\vspace{-0.5cm}
\end{table}

The four decay modes of table \ref{tab:rare} can probe different classes of NP; 
those with $\ell^+ \ell^-$ 
in the final state are affected by irreducibile backgrounds, and their 
SM contributions include large-distance parts which are under worse theoretical 
control.
The modes with neutrinos in the final state are cleaner, and represent the 
new ``holy grail'' of kaon physics. 
Precise measurement of these decays offers the potential to clarify the flavour 
structure of any NP directly observed in high-energy colliders, or to indirectly 
probe flavour effects in NP up to scales of order 100 TeV.

\begin{figure}[hbt]
\centering
\includegraphics[width=\linewidth]{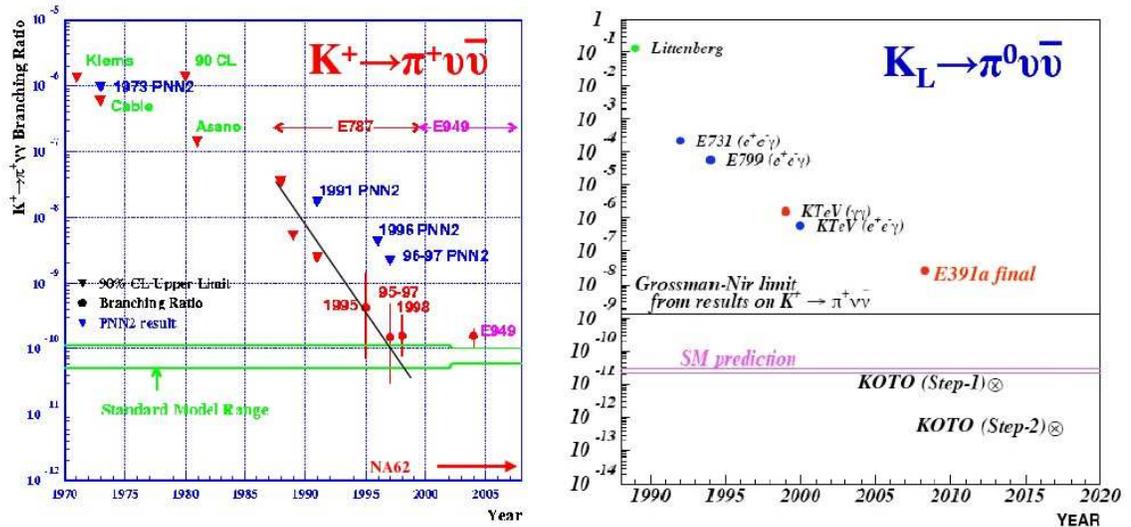}
\caption{Experimental progress on the measurement of $K \rightarrow \pi \nu 
\overline{\nu}$ BRs.}
\label{fig:Kpnn}
\end{figure}

Experimental detection of $K \rightarrow \pi \nu \overline{\nu}$ decays presents 
formidable difficulties, due to the $10^{10}$ background/signal ratio coupled with 
non-closed kinematics due to undetected neutrinos.
Seven $K^+ \rightarrow \pi^+ \nu \overline{\nu}$ candidates were detected in the 
dedicated E787 and E949 BNL experiments \cite{E787-E949} using low-energy kaon 
decays at rest. Only limits exist for $K_L \rightarrow \pi^0 \nu \overline{\nu}$, 
some 3 orders of magnitude higher than the SM prediction \cite{PDG}; see figure 
\ref{fig:Kpnn}.

A program aimed at the measurement of $K_L \rightarrow \pi^0 \nu \overline{\nu}$ 
is being pursued in Japan. The concept involves an intense thin ``pencil'' neutral 
beam entering a photon-hermetic vacuum decay region. The pilot project E391a at 
KEK took data in 2004 and 2005 with $\sim$ 2 GeV/$c$ $K_L$ to test the experimental 
approach, and significantly improved the BR upper limit \cite{E391a}.
The first experiment aiming to a measurement is KOTO, presently in the 
commissioning phase at J-PARC \cite{KOTO}. The optimization of the new neutral 
beam line, and the better duty cycle and acceptance lead to an estimate of 2.7 
SM events with 2.2 background events in 3 years, with a first run in 2011. 
A second phase is planned, involving a dedicated beam line and an improved detector, 
aiming at a O(100) event measurement.

Improvement on $K^+ \rightarrow \pi^+ \nu \overline{\nu}$ is expected from the 
NA62 experiment at CERN \cite{NA62}. This uses a new decay-in-flight approach 
with 75 GeV/$c$ kaons in an unseparated beam and kinematic selection of 
slow pions which, combined with photon quasi-hermeticity and particle ID, lead  
to the expectation for 55 SM events per year with $\sim$ 20\% background.
The experiment completed the R\&D phase and started construction, aiming for 
a first test run in 2012.

\vspace{-0.5cm}
\section{Exciting times}

A long-standing strong interest in $K \rightarrow \pi \nu \overline{\nu}$ materialized 
in the US as a number of interesting proposals which, unfortunately, could not be 
funded \cite{US}. Fermilab's plans on high-intensity physics involve kaon decays 
as a central topic, and the physics goals of a proposed future 2 MW proton source 
(``Project X'') \cite{ProjectX} include ``ultimate precision'' rare kaon 
experiments, aiming for $\sim$ 500 (200) $K^+ \rightarrow \pi^+ \nu \overline{\nu}$ 
($K_L \rightarrow \pi^0 \nu \overline{\nu}$) decays per year with signal to 
background above 4. 
Such a facility would also allow an improved CPT test on $m(K^0)-m(\overline{K}^0)$, 
sensitive to effects at the Planck scale. Construction could start in 2015, 
with first physics in 2020.

Limited space precludes the description of many other interesting recent results 
from $K$ decays, as well as a detailed discussion of ongoing programs such as the continuation of 
KLOE-2 \cite{KLOE2}, aiming at an eightfold increase in statistics.
Finally, the TREK project in preparation at J-PARC \cite{TREK} should be mentioned, 
aiming at a new search for non-SM T-violating transverse muon polarization in 
$K^+_{\mu 3}$ decays, with a tenfold improvement over the limit from the previous 
KEK experiment \cite{E246}.

\vspace{-0.5cm}
\section{Bright conclusions}

This brief review gives a bird's eye view of some of the fundamental physics 
which can be studied with kaons; detailed results can be found in other 
contributions to these proceedings and in the PDG compilation \cite{PDG}.
Kaons provided fundamental input to the building and the quantitative testing 
of the SM. They now enter as prime actors in the present era of precision 
tests, through challenging ultra-rare decay probes offering very powerful 
opportunities, pursued in many different laboratories.
It is guaranteed that these ``minimal'' probes of flavour effects will play 
a prominent role also in future editions of the CKM conference.

\end{document}